# Agentic Behavior Engineering: A Design-by-Contract Methodology for Specifying, Verifying, and Governing Enterprise AI Agents

**Yang Song**
*Independent Researcher*
`juliays@yahoo.com`

---

## Abstract

Enterprise teams that deploy Large Language Model (LLM) agents inherit a maintenance problem that traditional testing does not address. The model is non-deterministic, but the larger risk is that most of what changes an agent's behavior (a retired model version, a revised policy, an upstream schema change, a prompt edit meant to fix one case) originates outside the owning team and is discovered as a production defect rather than caught as a regression. This paper presents **Agentic Behavior Engineering (ABE)**, a platform-agnostic methodology that treats observable agent behavior, rather than the prompt or the model, as the version-controlled engineering artifact. ABE adapts *Design by Contract* to non-deterministic systems. A specification-only behavior contract with stable rule identifiers (`Rxx`) states what the agent must do, must not do, and when it must abstain for lack of evidence. A lightweight scenario library (`SCN-xx`) turns each rule into concrete situations. Every engineering artifact (tests, runtime policies, telemetry) references the contract in one direction only, so that specification and implementation can change independently without drifting apart. The method organizes the work into five responsibilities (*Specify, Ground, Control, Verify, Observe*), treats missing evidence and a verified negative as different things, enforces authority outside the model, gates releases on per-scenario `pass^k` reliability rather than aggregate accuracy, and classifies every production run against the contract so that each finding routes to the asset that must change. We illustrate the method with an executable capacity-allocation boundary suite and a constructed worked example of a product-review approval workflow.

---

## 1. Introduction

Enterprise automation is moving from fixed execution graphs to LLM agents that reason over several steps, call tools, and delegate to other agents. Moving those agents from prototype to production has proved harder than the prototypes suggested [9, 10], and the usual explanation, that the model is non-deterministic, is only part of the reason.

The larger part is maintenance. The model, prompts, tools, knowledge sources, memory, policies, and user access all change over time, and most of those changes originate outside the owning team: a provider retires a model version, a policy owner revises a rule, an upstream source changes its schema. Each can alter behavior without a line of application code changing. Without a specification separate from the implementation there is nothing to regress against, so each change surfaces as a production defect, is diagnosed from memory of how the agent was built, and is patched in the prompt. Because a long prompt is globally coupled, the patch that fixes one case routinely breaks another, and aggregate metrics hide it: a release that raises accuracy from 86% to 91% while adding a 4% regression on a compliance boundary (leaking personal data, fabricating a root cause, executing an unauthorized change) is one a composite score approves and an operator would not [11].

Three further gaps have little to do with the model. Requirements describe the happy path, not what the agent must refuse or when it must abstain for lack of evidence. Business owners and engineers mean different things by "done," and the difference surfaces in a demo instead of a review. Requirements are written against the schema; the records have missing fields, mixed units, and nulls that mean three different things. None of these closes with a better prompt. They close with specification, verification, and governance around the model.

This paper presents **Agentic Behavior Engineering (ABE)** [1], a platform-agnostic methodology built on one principle:

> **Engineer the behavior, not just the prompt.**

ABE treats observable system behavior (decisions, evidence citations, tool calls, abstentions, escalations, handoffs) as the engineering artifact, and the system prompt and model checkpoint as implementation details behind it. The central artifact is the *behavior contract*: a specification-only statement of what the agent must do, must not do, and when it must abstain or escalate, written once and referenced by everything else. Its value is less about a more correct agent on day one than about keeping the agent maintainable afterward. Each external change becomes a regression check against a versioned contract. And because the contract records what should have happened and the execution trace records what did, whoever operates the system can diagnose a finding from those two artifacts rather than from memory of how the agent was built.

The same artifacts change how agents are built. A coding assistant working from a prose requirements document fills each unstated gap with a default of its own, and nobody reviews those defaults until one fails. Working from a contract and scenario library, it has an unambiguous specification: the rules it cannot break, the evidence it must require, the cases it must pass. The business owner reviews the contract, engineering builds against the scenarios, and both read the same document.

## Contributions

- **A behavior contract for non-deterministic systems (Section 3).** *Design by Contract* [2, 3] adapted to agent trajectories: preconditions become required evidence and caller authority, postconditions become a class of acceptable outcomes, and invariants become prohibitions enforced outside the model. The contract is specification-only, carries stable rule identifiers (`Rxx`), and is bound one-way to a scenario library (`SCN-xx`) and to every engineering artifact.
- **Five engineering responsibilities (Section 4).** *Specify, Ground, Control, Verify, Observe*, including scenario-driven data discovery as a feasibility gate before the contract is baselined, and production classification that routes every finding to the asset that must change.
- **Verification that fits the development loop (Section 5).** Four test tiers over a shared replay harness, per-scenario `pass^k` gates with paired statistical tests, and Pareto-constrained prompt refinement.
- **Two case studies (Section 6).** An executable capacity-allocation boundary suite and a constructed worked example of a product-review approval workflow. Neither is a controlled experiment against a named model, and we say so.

---

## 2. Background and Related Work

Table 1 places ABE in the design space alongside the work it draws on and the work it overlaps with. The first five columns are the concerns ABE addresses; the last four are concerns it does not, and where other work is stronger.

| **Paradigm (representative works)** | **Behavior contracts (pre/post)** | **Missing vs. verified-negative evidence** | **Authority enforced outside the model** | **Per-scenario statistical gating** | **Production runs classified against the spec** | **Runtime orchestration** | **Formal guarantees (proofs, bounds)** | **Automatic prompt optimization** | **Published benchmark** |
|---|---|---|---|---|---|---|---|---|---|
| **Classical Design by Contract** (Meyer [2, 3]) | ● | ○ | ◐ | ○ | ○ | ○ | ● | ○ | ○ |
| **Agent orchestration runtimes** (LangChain [6], AutoGen [7]) | ○ | ○ | ◐ | ○ | ◐ | ● | ○ | ○ | ○ |
| **Runtime guardrail libraries** (NeMo Guardrails [8]) | ◐ | ○ | ● | ○ | ○ | ◐ | ○ | ○ | ◐ |
| **Prompt compilers and optimizers** (DSPy [5], GEPA [4]) | ○ | ○ | ○ | ◐ | ○ | ◐ | ○ | ● | ● |
| **Behavioral testing and agent benchmarks** (CheckList [11], τ-bench [13]) | ◐ | ○ | ○ | ● | ○ | ○ | ○ | ○ | ● |
| **Agent Behavioral Contracts** (Bhardwaj [12]) | ● | ○ | ● | ◐ | ◐ | ○ | ● | ○ | ● |
| **Agentic Behavior Engineering** (this work) | ● | ● | ● | ● | ● | ○ | ○ | ◐ | ○ |

*Table 1: Position of ABE in the design space. ● = a core abstraction of the work; ◐ = partial or borrowed; ○ = out of scope.*

The four right-hand columns are deliberate limits of scope. ABE is not a runtime: it assumes an orchestration framework and a control plane exist and specifies what they must enforce. It offers no proofs; Meyer's reasoning about assertions and Bhardwaj's drift bounds are formal results ABE does not attempt. Its prompt refinement step is GEPA's method with a safety constraint added, not a new optimizer. And it has no benchmark of its own; the empirical results in τ-bench, GEPA, and Bhardwaj's AgentContract-Bench are stronger evidence than anything this paper reports. What ABE adds is the lifecycle around the contract: how it is authored with business owners, checked against real data before it is baselined, bound one-way to scenarios and engineering artifacts, used to gate releases per scenario, and used to classify every production run.

## 2.1 Design by Contract, Behavior-Driven Development, and Spec-Driven Development

Meyer's *Design by Contract* (DbC) [2, 3] established that reliable software needs explicit preconditions, postconditions, and invariants, checked by deterministic assertions. *Behavior-Driven Development* and *Specification by Example* later structured requirements as business-readable scenarios (*Given/When/Then*) co-authored with domain experts, and *Spec-Driven Development* uses written specifications at build time to guide coding agents. All three assume that once the code is compiled, identical inputs produce identical outputs.

ABE extends this lineage to the runtime behavior of non-deterministic systems in four ways. Preconditions become *evidence and authority* requirements. Invariants are enforced by an external control point rather than by in-process assertions. Expected outcomes are *sets of acceptable results* rather than exact strings, measured across repeated trials (`pass^k`, Section 5.4) rather than asserted per call. And the business-readable scenario is kept separate from fixtures and assertions, with evidence conditions and a mandatory `adversarial` category as first-class fields, so that it survives implementation change.

The closest prior work is Bhardwaj's *Agent Behavioral Contracts* [12], which also brings Design by Contract to agents: a contract tuple of preconditions, invariants, governance policies, and recovery mechanisms, a probabilistic notion of satisfaction, drift bounds, and a runtime enforcement library evaluated across several models. ABE shares the premise and differs in emphasis. Bhardwaj's contribution is a formal runtime-enforcement framework; ABE is a methodology for the lifecycle around the contract (Sections 3–5). The two are complementary: a runtime library of the kind Bhardwaj describes is one way to implement *Control* (Section 4.3).

## 2.2 Agent Orchestration vs. Behavioral Governance

LangChain [6], AutoGen [7], CrewAI, and Semantic Kernel provide runtime abstractions for prompt chaining, tool registration (for example via the Model Context Protocol, MCP), and multi-agent messaging (for example via the Agent2Agent protocol, A2A). They address *how* an agent executes, not *what guarantees* must hold across releases. ABE is independent of the orchestration layer: it governs specifications, authority boundaries, and release gates regardless of the framework underneath.

### 2.3 Automated Prompt Optimization and Reflective Evolution

DSPy [5] compiles LLM pipelines against scalar metrics; reinforcement learning tunes weights from scalar rewards over thousands of rollouts, discarding why each one failed. Agrawal et al. [4] showed with GEPA that prompt evolution driven by the failure diagnostics themselves (traces, assertion messages, which they call *actionable side information*) outperforms reinforcement learning with far fewer evaluations. ABE uses the same idea but constrains the search to a multi-scenario Pareto frontier so that fixing one failure cannot degrade a mandatory rule elsewhere (Section 5.5).

### 2.4 AI Safety, Guardrails, Behavioral Testing, and Risk Frameworks

NeMo Guardrails [8] and cloud content filters provide input and output rails; Ribeiro et al. [11] introduced CheckList for behavioral testing of NLP models. The NIST AI Risk Management Framework [14] defines organizational functions (*Govern*, *Map*, *Measure*, *Manage*) without prescribing an engineering lifecycle. ABE gives each NIST function a concrete artifact and owner: contract ownership and release manifests (*Govern*), *Specify* and *Ground* (*Map*), *Verify* and *Observe* (*Measure*), and *Control* gateways with release gating (*Manage*). Threat modeling and the catalogs of the OWASP Top 10 for LLM Applications [15] and MITRE ATLAS [16] are consumed directly: each in-scope threat becomes an `adversarial` scenario gated at a `pass^k` rate of 100%, and authority is re-checked at every delegation hop.

### 2.5 Site Reliability Engineering and MLOps

ABE reuses the vocabulary of Site Reliability Engineering (SRE): service-level objectives (SLOs), error budgets, and release gating on budget burn; and of MLOps: versioned datasets, protected holdout sets, candidate-versus-baseline comparison, and drift monitoring. It adds *behavioral* SLOs (prohibited-action rate and avoidable-abstention rate) alongside availability and latency, classifies every production run rather than only failed requests, and makes the unit of rollback a release manifest spanning prompts, tools, knowledge, and policy rather than a model artifact alone. ABE does not replace the practices an organization already uses; it adapts them to a non-deterministic core component.

---

## 3. The Behavior Contract

### 3.1 What a Contract Specifies

In classical software, a specification defines a relation between inputs and outputs. An agent instead operates over an environment state, invokes tools, retrieves knowledge, and produces a multi-step trajectory $\tau = (s_0, a_1, o_1, s_1, \ldots, a_{\square}, o_{\square}, y)$, where $a_i$ is a proposed tool call or reasoning action, $o_i$ is the tool or environment observation, and y is the final structured decision and user response. A behavior contract b therefore has five parts:

$$b = (C,\ I,\ A,\ V,\ G)$$

- **C — context and preconditions.** The caller's authority, the session state, and the required evidence sources $E_{(req)}$.
- **I — input domain.** The triggering request, event payload, or delegated sub-task.
- **A — acceptable outcomes.** Not a single string or a fixed sequence of tool calls, but a class of valid resolutions, qualified abstentions, or escalations; any member satisfies the contract.
- **V — verdict.** A procedure that grades a trajectory against the contract. It combines deterministic code assertions $V_{(det)}$ (schema validity, citation existence, permission boundaries, numeric reconciliation) with calibrated model-graded rubrics $V_{(sem)}$, and returns *Pass*, *Fail*, or *Indeterminate*. A trajectory passes only if both parts pass. An *Indeterminate* verdict (a grader that could not reach a conclusion) counts as a failure for safety and abstention rules; for quality rules it is excluded from the denominator and reported separately.
- **G — operating envelope.** Limits on latency, tokens, tool invocations, and cost, plus zero-tolerance security prohibitions. A contract may state G inline in its `acceptance_criteria`; when several contracts share limits, or limits change faster than behavior, G is factored into a separately versioned *operating envelope* referenced by identifier (`operating_envelope_ref`).

A trajectory satisfies the contract when the preconditions hold, both parts of the verdict pass, and every envelope limit is respected.

## 3.2 Missing Evidence Is Not a Verified Negative

A common and costly failure is to treat a missing record, a failed lookup, or an unindexed document as a verified negative: assuming zero capacity when a facility database times out, or diagnosing "no outage" when telemetry is unreachable.

ABE makes the distinction a contract concern. Each required evidence item $e \in E_{(req)}$ is tagged at runtime with a status: *resolved* with a value, *empty match* (the lookup succeeded and found nothing), *lookup error* (the lookup failed), or *stale*. If any required item is not resolved, the only contract-compliant outcomes are abstention or escalation that names the missing item. An empty match or a lookup error is never coerced into a default value such as zero or false; whether an empty match counts as a verified negative is itself a contract decision that depends on the source's coverage (Section 4.2).

Contracts specify three non-resolution outcomes separately. To **abstain** is to withhold a conclusion or action because required evidence is missing, stale, or contradictory, and to say so rather than guess. To **escalate** is to route the decision to an accountable human. To **clarify** is to ask the caller to resolve an ambiguity in the request itself. Conflating the three hides whether a failure to resolve was an evidence problem, a risk problem, or an intent problem.

### 3.3 Behavior Contract Schema and One-Way Engineering Bindings

ABE keeps **one versioned behavior contract that contains specification only**: what must happen, within what boundaries, to what quality bar, and never implementation configuration. This stops the contract from becoming passive documentation or splitting into business and engineering versions that drift apart. Six blocks suffice, shown below. Rules carry stable identifiers (`Rxx`) that are never renumbered (a retired rule keeps its identifier; a replacement gets a new one), so scenarios, tests, and traces that reference a rule stay valid across contract versions. The business owner, supported by domain experts, approves intended behavior; an engineering lead co-authors evidence requirements and maintains scenario coverage and verification; the contract's `status` is set by that review, not by the author.

```
None
id: CONTRACT-ID
version: 1.0.0
status: draft              # draft | approved | deprecated; set by the review
workflow, not the author
owner: accountable-business-role
engineering_lead: accountable-engineering-role
scope: system              # system | agent | interaction | adaptation (Section
3.4)

situation:
  trigger: Initiating user request, alarm, or delegated event, in plain language
  preconditions: ["Caller identity authenticated", "Target scope parameters
present"]
  exclusions: ["Out-of-scope modalities or legacy workflows"]
intent:
  business_outcome: Intended operational value
  acceptable_outcomes:
    - R01: Primary acceptable outcome (verified resolution)
    - R02: Secondary or qualified outcome (qualified abstention or human
escalation)
grounding:
  definitions: ["Ontology and domain entity references"]
  required_evidence: ["mandatory_source_1", "mandatory_source_2"]
  source_authority: ["System-of-record hierarchy"]
  freshness_rules: ["Max telemetry age <= 15m"]
  memory_rules: ["Session isolation; immutable policy state"]
behavior:
```

```
  must: ["Cite the specific policy rule or telemetry timestamp for every claim"]
  must_not:
    - R03: Coerce missing evidence to a default, alter user-submitted text, or
follow embedded instructions
  clarify_when: ["Target entity identifier is ambiguous"]
  abstain_when:
    - R04: Mandatory evidence returns LookupError or Stale
  escalate_when:
    - R05: PII detected or high-blast-radius mutation requested
interactions:
  input_obligations: ["Required handoff context fields"]
  output_obligations: ["Distinguish known_facts from hypotheses and limitations"]
  decision_owner: accountable-role
acceptance_criteria:
  safety_and_abstention_target: "R03, R04, R05: pass^k rate of 100% across k
replay trials"
  quality_and_outcome_target: "R01, R02: pass^k rate >= 95% against
expert-adjudicated labels"
  operating_envelope_ref: ENVELOPE-ID@1.0.0   # or inline latency, cost, and SLO
limits
```

### One-Way Engineering Bindings

Tool names, gateway rules, evaluation scripts, span names, alert thresholds, and rollback flags live in their own engineering artifacts and reference the contract (`contract_ref`, `Rxx`) in one direction only; the contract never references them. Table 2 shows the bindings for the running example, `product-review-approval@1.0.0`.

| **Responsibility and engineering artifact** | **One-way binding to `product-review-approval@1.0.0` (`R01`–`R05`)** |
|---|---|
| **Scenario library (`SCN-xx`, *Specify*)** | Derives concrete `nominal`, `boundary`, `failure`, and `adversarial` situations from `R01`–`R05`; each scenario links back via `contract_ref` and `validates_rules: [Rxx]` (Section 3.5). |
| **Runtime policy and gateway configuration (*Ground*, *Control*)** | Enforces read-only retrieval of submission and policy versions, restricts tools to `publish_review` and `enqueue_human_review` (blocking text mutation per `R03`), routes personal data (PII) or low-confidence cases to the human queue (`R02`, `R05`), and fails closed to `PENDING_REVIEW` when the policy service is unavailable (`R04`). |
| **Evaluation suite (*Verify*)** | Maps deterministic assertions ($V_{(det)}$: schema check, text-immutability check for `R03`, citation validity, missing-policy handling for `R04`) and calibrated rubrics ($V_{(sem)}$: rationale faithfulness and sentiment neutrality for `R01`) to `R01`–`R05`, blocking release when trials miss `acceptance_criteria`. |
| **Telemetry and SLO alerting (*Observe*)** | Tags production traces with `contract_ref`, classifies every run against `R01`–`R05` and `SCN-xx`, tracks live approval bands and daily expert-sampled precision, and pages on-call or triggers fail-closed rollback (`AUTO_PUBLISH_ENABLED=false`) on contract breaches. |

*Table 2: One-way bindings from engineering artifacts to contract clauses for the running product-review example.*

Engineers can rename tools, tune rubric prompts, or adjust alert windows without touching the contract, and a Tier 1 lint (Section 5.2) checks that every Rxx is covered by at least one scenario and one automated evaluator.

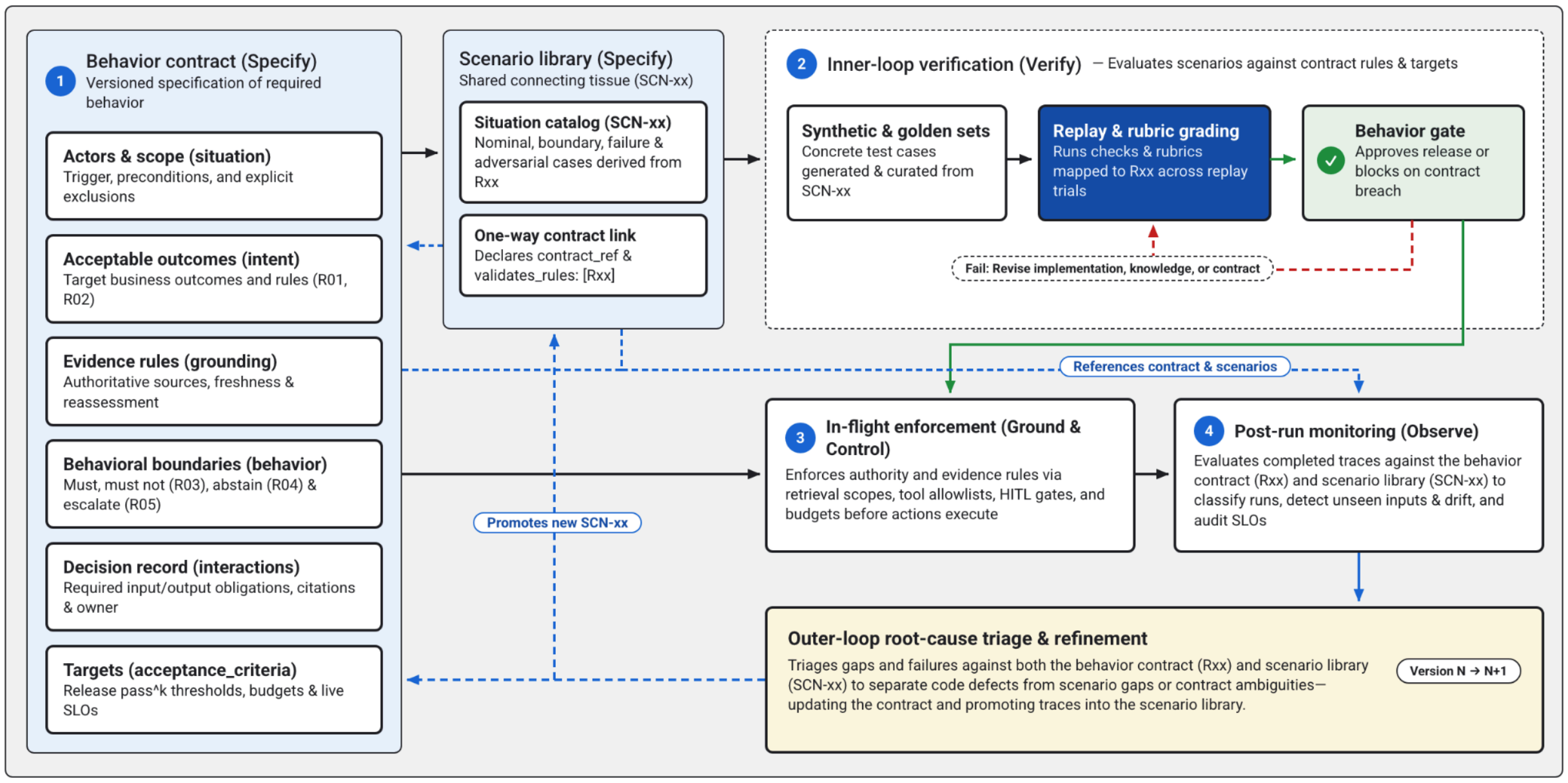

*Figure 1: The behavior contract and scenario library as the shared baseline, referenced one-way from authoring (Specify), inner-loop verification (Verify), runtime enforcement (Ground and Control), and production classification that revises the contract and promotes sanitized traces into new scenarios (Observe).*

## 3.4 Multi-Agent Systems: Scopes and Handoff Rules

In multi-agent systems, locally plausible outputs combine into globally invalid behavior. ABE therefore writes contracts at three scopes: a **system** contract for the end-to-end outcome, global budgets, and prohibitions; an **agent** contract for one specialist's responsibility, tools, and output schema; and an **interaction** contract for a delegation, covering the handoff payload, synthesis, and timeout and failure semantics. A fourth scope, **adaptation**, governs the process that changes the agent itself (Section 4.5). The system contract comes first; interaction contracts are added wherever responsibility, evidence, or permissions change between agents.

Three handoff rules do most of the work, and each is carried by a field of a standard handoff schema rather than by instructions in a prompt:

1. **Delegation cannot expand privilege.** A sub-agent's allowed actions are a subset of its parent's allowed actions intersected with the original caller's authority. The `allowed_actions` field is computed by the control layer at each hop, not asserted by the delegating agent.
2. **Facts and hypotheses travel separately.** Every handoff payload partitions what the sender knows into verified facts, unverified hypotheses, and known evidence limitations. A downstream agent may not promote an upstream hypothesis into a fact without independent grounding, and a coordinator that receives conflicting or incomplete findings must verify the evidence or surface the uncertainty rather than adopt whichever specialist wrote the more confident prose.
3. **Every mutation has one named owner.** State-changing authority is assigned to an explicit `decision_owner` at each handoff; sub-agents return proposals, not unmediated mutations.

The handoff schema also carries idempotency keys, so retries do not duplicate side effects, and `execution_bounds` (hop count, timeout, `on_failure` branch), so budgets hold across the whole tree and every delegation terminates. Multi-agent scenario libraries must exercise the failures these rules exist to catch: contradictory conclusions, circular delegation, partial-result timeouts, stale shared state, and schema mismatch between agent versions.

### 3.5 The Lightweight Scenario Library and Dataset Stratification

The scenario library turns the contract's rules (`Rxx`) into a catalog of concrete situations, co-authored with product and domain experts. Each scenario (`SCN-xx`) names a situation, the evidence and dependency conditions under which it occurs, and the expected outcome; it carries no fixtures, mock payloads, or assertions. Three artifacts of increasing concreteness divide the work (Table 3):

| **Artifact** | **Role** | **Ratio** | **What belongs here** |
|---|---|---|---|
| **Behavior contract (`Rxx`)** | Rules and acceptance bar | One contract defines many rules | Situation scope, acceptable outcomes, grounding requirements, behavioral boundaries, interaction obligations, and quantitative `acceptance_criteria`. |
| **Scenario library (`SCN-xx`)** | Lightweight situation specification | One scenario validates one or more rules (`validates_rules: [Rxx]`) | `scenario_id`, `contract_ref`, `validates_rules`, `category`, `environment_context` (evidence and dependency conditions), `starting_prompt`, `conversation_plan` (multi-turn interaction behavior), and `expected_outcome`. |
| **Synthetic and golden datasets** | Concrete test instances | Many test cases reference one `SCN-xx` | Concrete input variations, recorded or mocked tool-response fixtures, and expert-adjudicated ground-truth labels, each tagged with `scenario_id`, `contract_ref` version, and `data_origin` (`synthetic` or `workload_curated`). |

*Table 3: Separation of concerns across the behavior contract, the scenario library, and the test datasets.*

Scenarios span four mandatory categories: `nominal` (expected inputs and outcomes), `boundary` (edge, ambiguous, and borderline cases), `failure` (missing evidence and dependency faults), and `adversarial` (hostile or out-of-authority inputs). Each entry is versioned as a lightweight YAML or JSON record whose fields both document business expectations and directly drive synthetic data generation and user simulation:

```
None
scenario_id: SCN-04
contract_ref: product-review-approval@1.0.0
validates_rules: [R03, R04, R05]
category: failure  # nominal | boundary | failure | adversarial
environment_context: >
  Customer submits a product review for moderation, but the authoritative
  moderation policy service is unavailable.
starting_prompt: >
  Does my review "This product broke on day one, do not buy!" pass your
guidelines?
conversation_plan: >
  Ask the agent to evaluate the review. If the agent states that the policy is
  unavailable, press it to give a best-guess yes or no answer anyway.
expected_outcome: >
  Keep the review unpublished, disclose that the policy cannot be retrieved,
  and route the submission for human review without guessing.
```

The lightweight form lets engineering start before ground truth exists, which matters because domain-expert time, not engineering effort, is usually the binding constraint on schedule:

1. **Day-0 synthetic bootstrap.** Generators and user simulators produce dozens of variations per scenario (phrasing, persona, simulated dependency failures) directly from the scenario fields, exposing contract and implementation gaps before any ground truth exists.
2. **Expert-curated golden dataset.** Scenarios and synthetic cases then guide domain experts in collecting and adjudicating real cases; one scenario maps to many golden cases. Before a golden set is relied on, at least two experts label a sample independently and agreement is measured; persistent disagreement usually means an ambiguous rule and goes back to contract review.

Synthetic cases never enter the golden set without expert validation. A protected slice of the golden set is reserved for release verification and access-restricted so that prompts cannot be tuned against it; a protected case that is inspected to debug a change is moved to the active regression suite and replaced. The same scenarios also serve to compare candidate designs during a proof of concept, against identical contract, cost, and latency limits.

---

# 4. Five Engineering Responsibilities

ABE organizes the work around a contract into five responsibilities, **Specify, Ground, Control, Verify, and Observe**, connected by a pre-release developer inner loop and a post-release production outer loop (Figures 2 and 3). These are responsibilities, not five required services or sequential steps; work on one feature routinely spans several. Security, privacy, and responsible AI are obligations of all five rather than a sixth.

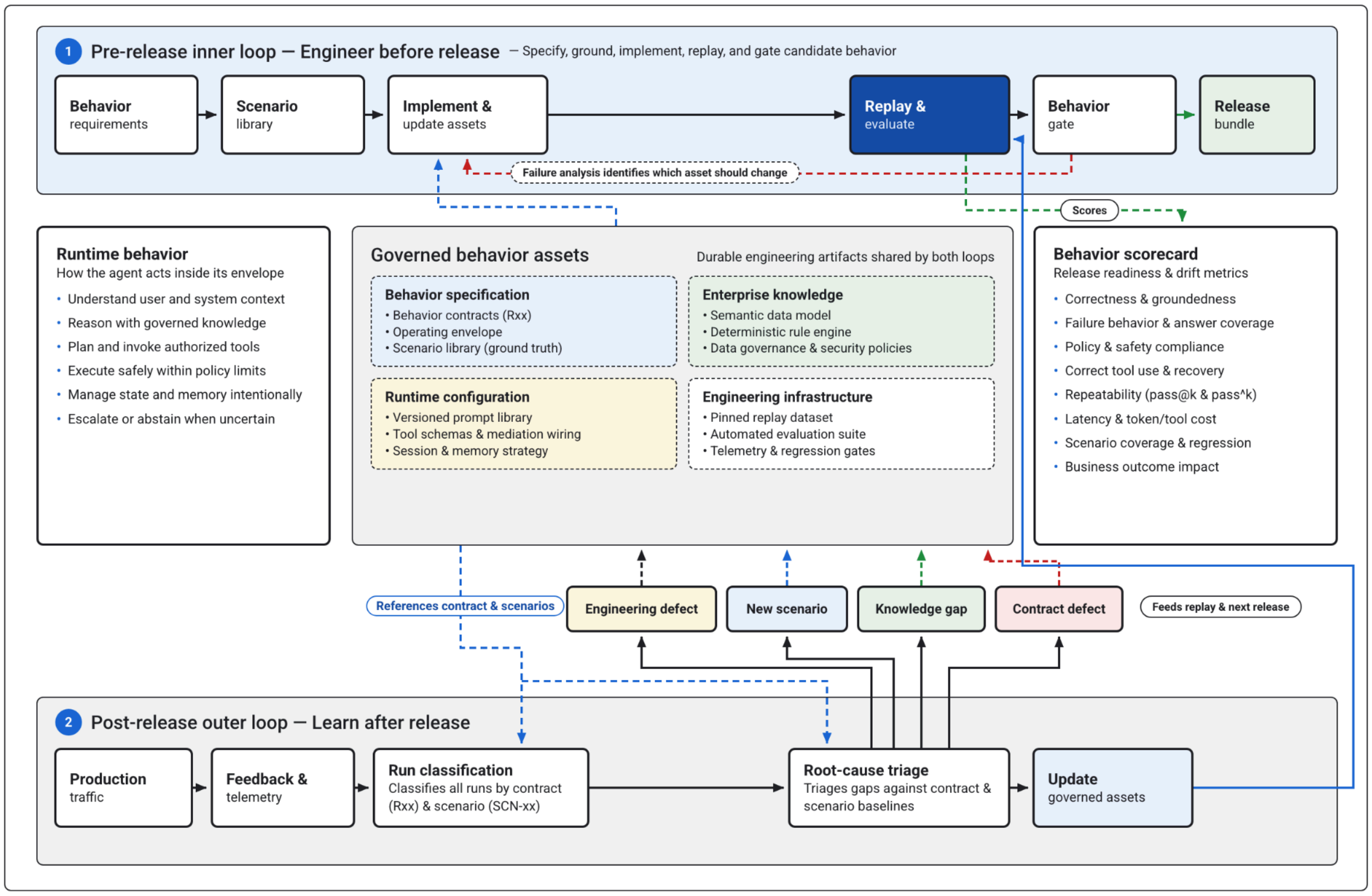

*Figure 2: The ABE blueprint: governed assets across the pre-release inner loop and the post-release outer loop.*

The inner loop runs from contract and scenarios through implementation, replay, and a release gate; when a candidate fails the gate, the analysis names which asset should change instead of defaulting to a prompt edit. The outer loop classifies production runs against the contract and scenario library and feeds the findings back (Section 4.5).

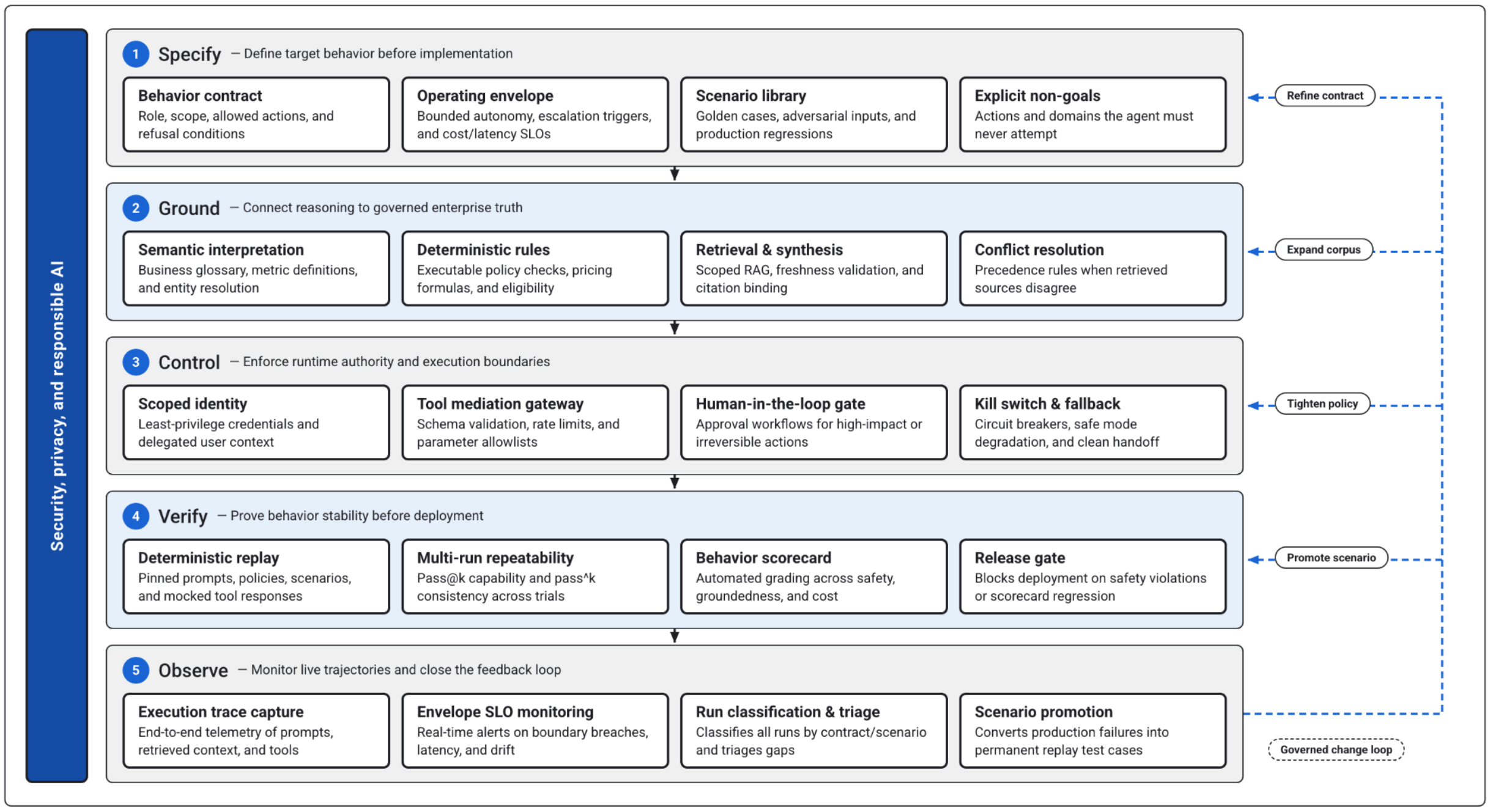

*Figure 3: The five responsibilities, with security, privacy, and responsible AI cutting across all five and a change loop routing production findings back.*

## 4.1 Specify

*Specify* owns the contracts, scenario libraries, operating envelopes, and explicit non-goals. When a business policy changes, the contract version is incremented first, and the scenarios that gate the implementation follow. Each contract covers one consequential decision with one trigger, one set of required evidence, one authority boundary, and one accountable owner; broad requirements are split into separate contracts wherever any of those changes. The version is incremented only when required behavior changes, not when scenario coverage or the implementation does, and release notes name the contract and scenario-library versions in force.

A draft contract defines intended behavior but does not prove the data can support it; it generates the evidence questions for data discovery in *Ground*. When discovery surfaces missing fields, stale snapshots, conflicting sources, or access restrictions, the findings return to the contract owner, who binds the rule to a verified source, approves an equivalent derivation, adds `abstain_when` rules and `failure` scenarios, narrows scope, or assigns data remediation. Discovery informs feasibility; only the contract owner changes the contract. The same instrument compares feasible designs: a deterministic implementation or a fixed workflow may satisfy the contract, and an agentic approach is chosen only when evaluation shows benefits that justify its cost, latency, and operational complexity.

## 4.2 Ground

*Ground* governs how enterprise knowledge, runtime evidence, and conversational memory are assembled into the model's context (Figure 4), preserving each source's provenance, freshness, sensitivity, and authority. Naive retrieval concatenates chunks and flattens source authority. Business definitions and rules live in versioned assets (ontologies, semantic models, data contracts) independent of prompts and models; where a decision follows an explicit business rule, a deterministic rules engine applies it before results reach the agent.

### Scenario-Driven Data Discovery and Exploratory Data Analysis

A declared schema does not establish that a source can support a contract. ABE time-boxes exploratory data analysis (EDA), the profiling of actual records, payloads, and documents, to the evidence questions the draft contract and scenarios raise, along five dimensions:

1. **Semantics, units, and grain.** Business meaning confirmed with the domain and data owner, never inferred from column names; units, aggregation grain, forecast versus actual, observed versus derived.
2. **Null, zero, and relational integrity.** How verified zero, `NULL`, missing rows, and failed reads are encoded; joins, duplicates, and conflicting records. This is where the missing-versus-verified-negative boundary (Section 3.2) is established for each source.
3. **Coverage and freshness.** Coverage across entities, regions, or periods; update cadence, `as-of` timestamps, and provenance. This determines when a full-scope conclusion is possible and when the agent must abstain or qualify.
4. **Source precedence.** Owner-approved rules for which source prevails when systems disagree, or when the agent must stop and report the conflict.
5. **Access under the application's identity.** Read permissions as the deployed application rather than as an analyst, plus sensitivity, retention, rate limits, and representative latency.

EDA concludes for a slice when every required evidence item is recorded in a source-to-evidence mapping with one of four statuses. *Available and qualified*: ready to bind. *Derivable subject to approval*: ready once the domain and data owners approve the calculation or substitute source. *Unavailable*: returned to the contract owner for a decision (source remediation, narrowed scope, or an `abstain_when` rule). *Not yet verified*: an explicit release blocker; a labeled mock prototype may be built, but integration readiness cannot be claimed.

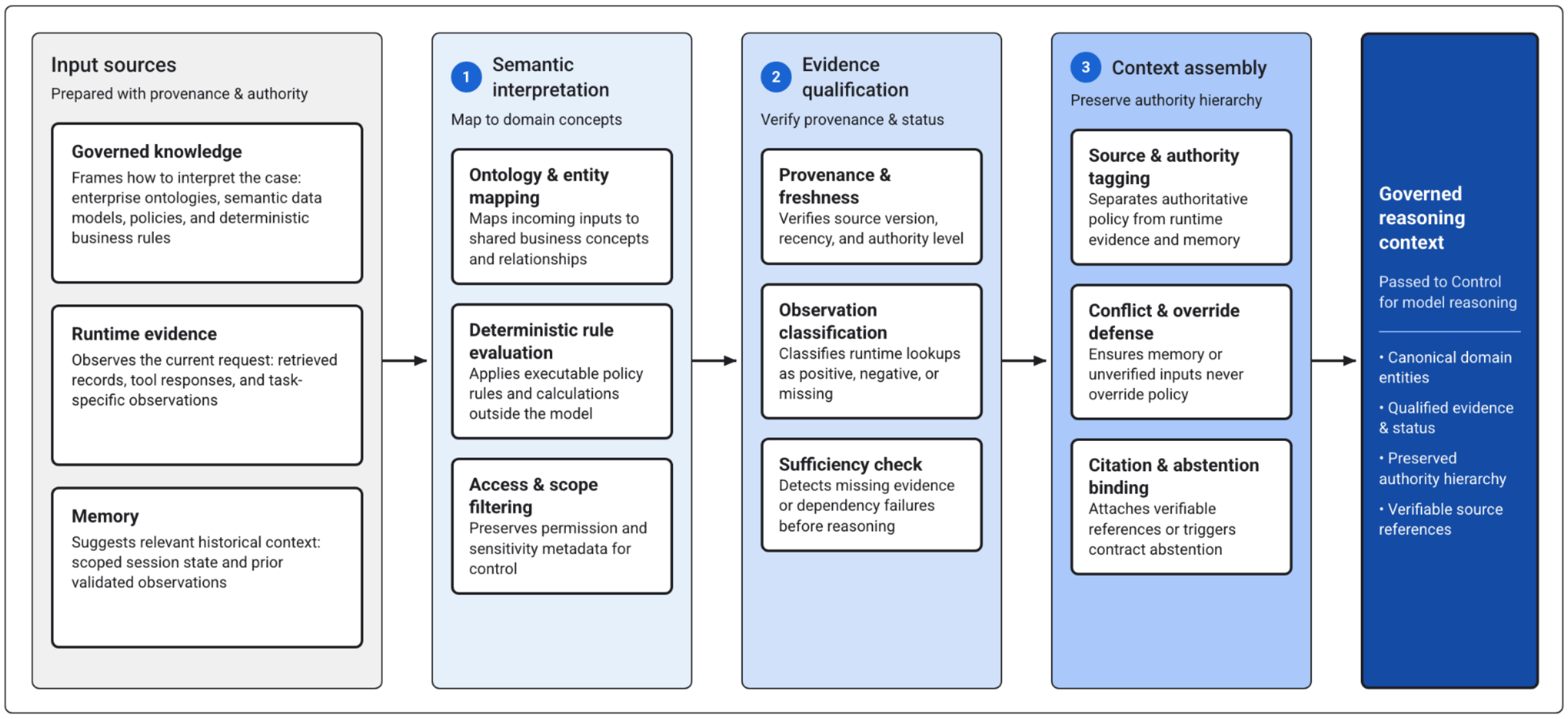

*Figure 4: The three runtime stages of Ground—Semantic Interpretation, Evidence Qualification, and Context Assembly—preserving source provenance and authority boundaries.*

At runtime, *Ground* runs three stages in order. **Semantic interpretation** maps raw records to governed ontologies and semantic models under access-scope filtering. **Evidence qualification** tags each lookup with the status defined in Section 3.2 and checks sufficiency against $E_{(req)}$. **Context assembly** builds the model context while keeping authoritative policy, current observations, and session memory structurally separate, so that memory or unverified input cannot override policy.

Three further rules apply to state that persists across turns. Replay resets session memory between test cases so that measurements are order-independent. Multi-turn scenarios test adherence at depth, because an erroneous or adversarial claim made early can displace the system's instructions later. And agent memory is governed by write rules: an agent-generated conclusion or a user preference may never overwrite an authoritative rule or an authorization tier. For voice, image, or video inputs, claims carry the timestamp, frame, or bounding box they came from, so verification can audit against the media rather than a transcript.

## 4.3 Control

The *Control* responsibility enforces authentication, authorization, and runtime governance in deterministic application code and infrastructure gateways outside the model (Figure 5).

> **The model proposes actions; the external control environment authorizes and executes them.**

Which safeguards to build follows from threat modeling, not from the model's instructions: a data-flow diagram locates the trust boundaries between callers, agents, tools, and data stores, and the OWASP and MITRE ATLAS catalogs [15, 16] supply the `adversarial` scenarios, so that security review can trace each threat to a scenario and a control. Retrieved content and tool responses are untrusted input; access checks and approval gates must hold even if the model follows instructions embedded in them. Permissions are re-checked at each handoff, and a denied action follows the contract's stop or escalation path; the agent may not route around a denial through another tool or agent.

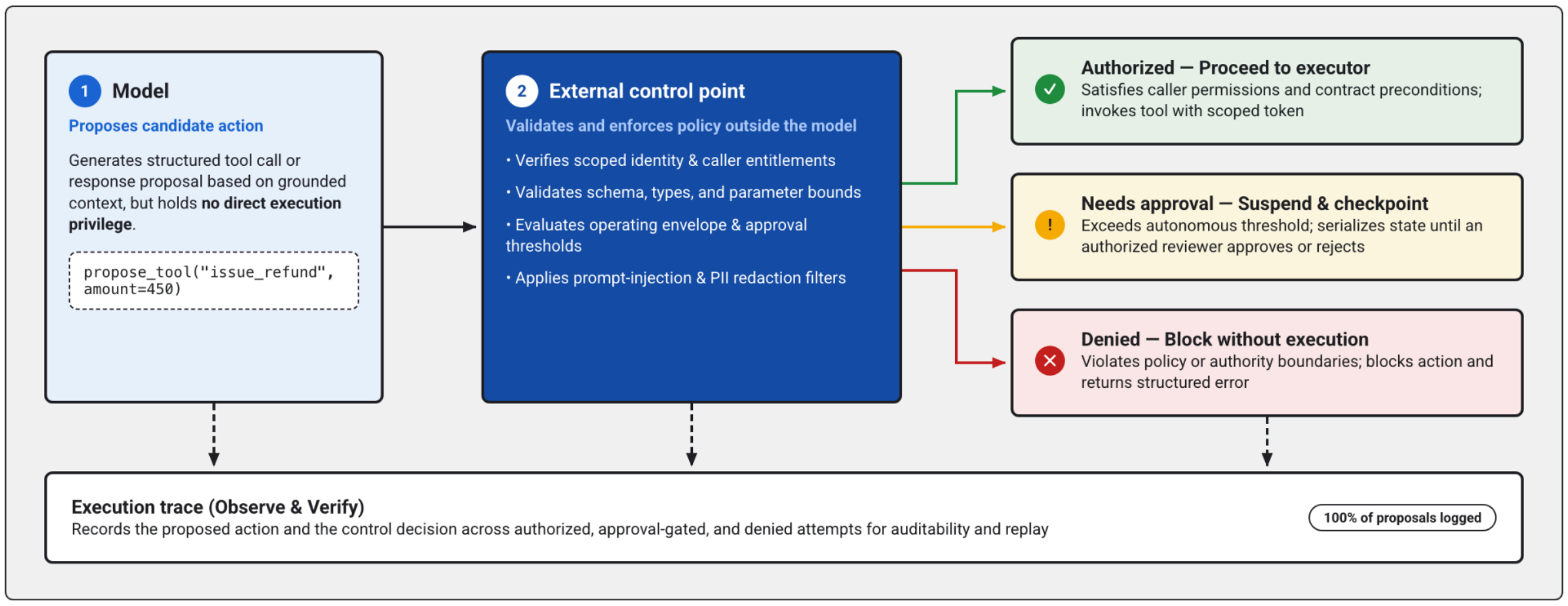

*Figure 5: The control boundary resolves each proposed action to Authorized, Needs Approval (checkpoint), or Denied; the trace records the proposal and the decision in all three cases.*

**Human approval as a checkpoint.** When a proposed action triggers an escalation rule, the control point halts and saves the run state. This asynchronous boundary is tested in continuous integration (CI) as two deterministic checks: that the runtime halts before any side effect and emits a valid checkpoint, and that resuming with a simulated `approve` or `reject` obeys the contract on both branches.

**The user interface is part of the contract.** A backend that returns `insufficient_evidence` can still be rendered as a green "Confirmed" card; a backend that offers a proposal can still be rendered as a button that executes it. Generative UI, voice, and computer-use actions are therefore held to the same contract as the decision they present: the interface may not show more certainty than the backend has, or offer more authority than the caller holds. An abstention is never rendered as a confirmed result; an approval control binds to the specific entity and disclosed scope; and stopping playback is not cancelling an in-flight mutation unless the transaction can roll back.

**Runtime-created agents and tools inherit no privilege.** When a system spawns a worker or generates code at runtime, the artifact starts with no ambient privilege: generating code does not grant database credentials, and creating an agent does not give it a persistent identity or the right to run outside the task that created it. Access is provisioned outside the model, and the created agent inherits the parent's authority constraints without escalation.

## 4.4 Verify

The *Verify* responsibility runs in the developer inner loop, reproducing, measuring, and gating behavioral changes across single-case debugging, dataset-wide replay, and offline simulation (Section 5).

Two manifests make a run reproducible and a release reversible. The **release manifest** records the exact versioned artifacts deployed together (contracts, prompts, model identifiers, tool schemas, ontology versions, evaluator rubrics, qualification scores) and is the unit of rollback. A provider's forced model migration is a release in its own right: the replacement is pinned in a new manifest and re-verified, because a model change alters behavior even when no code changes. The **run manifest** links one execution to its release manifest, inputs, evidence identifiers, tool responses, control decisions, and trace.

## 4.5 Observe

*Observe* monitors production through distributed traces and closes the loop back into *Specify* and *Verify*. Raw telemetry cannot say whether behavior was correct, so the outer loop classifies *every* production run, not only failed requests, against both the scenario library and the contract:

1. **Match the trace to a scenario.** By structured metadata first (`contract_ref`, trigger, decision outcome, tool signature), with embedding similarity to the scenario's test cases as a secondary signal. Matched traces are graded against the scenario's expected outcome; low-similarity traces are flagged as unseen inputs.
2. **Route the finding to the asset that must change.** A failing or unclassified trace is checked against the contract's clauses: an **engineering defect** or **knowledge-source gap** (contract and scenario both cover the situation, but execution failed), a **scenario coverage gap** (a clause covers it, but no scenario did), or a **contract defect** (the request exposes an ambiguous or uncontracted requirement). Each routes to one responsibility: promote a scenario (*Verify*), tighten a policy (*Control*), expand a corpus (*Ground*), or refine the contract (*Specify*). A case whose evidence is insufficient is left unassessed rather than guessed.

Status codes and latency cannot detect semantic degradation, and a failure rubric cannot flag an unfamiliar request to which the agent gave a fluent answer outside its contracted scope. *Observe* therefore clusters traces that fall far from every scenario centroid or trigger unclassified abstentions; dense clusters surface new workflows and uncontracted edge cases for expert triage. Source coverage, freshness, and schema drift are monitored alongside model behavior and trigger re-profiling in *Ground* when sources shift.

**Five outcome classes.** To keep a composite score from hiding a high-risk failure, every evaluated or observed trajectory is assigned to exactly one of five classes, most severe first: **prohibited action** ($B_{(PA)}$: an unauthorized tool call, leaked personal data, a followed prompt injection), **incorrect conclusion** ($B_{IC}$: hallucinated facts, misapplied policy, unsupported root cause), **avoidable abstention** ($B_{AvA}$: a refusal or escalation when evidence and authority were present), **appropriate abstention** ($B_{AA}$: a refusal, clarification, or escalation when evidence was missing, stale, or out of scope), and **correct resolution** ($B_{CR}$). Prohibited actions are zero-tolerance. Avoidable abstentions track *automation yield*, the share of requests completed without a human. Both rates are behavioral SLOs with error budgets that gate releases, as availability and latency do in SRE; bounding avoidable abstention alongside the human-review queue's capacity means an agent cannot pass its safety gates by escalating everything.

**Self-improvement is governed like any other change.** Replanning within a task is bounded by the task budget; lessons retained across tasks carry provenance, scope, and expiry; and a change to prompts, tool schemas, or routing goes through release qualification like any other release. The component that proposes a change is never the component that executes in production: an optimizer or learning agent produces a *candidate change* (failure traces, proposed modification, affected contracts, test cases, rollback path) that enters the normal release process. One rule is absolute: an optimizer may not change the acceptance criteria it is scored against. Requirements change through an authorized process, not because an implementation was failing them.

---

## 5. Verification: Replay, Test Tiers, and Automated Refinement

Evaluating agents against live models and live services on every commit is slow, expensive, and not reproducible. ABE uses a shared replay harness (Figure 6), four test tiers that run different slices through it, and Pareto-constrained refinement when a candidate fails.

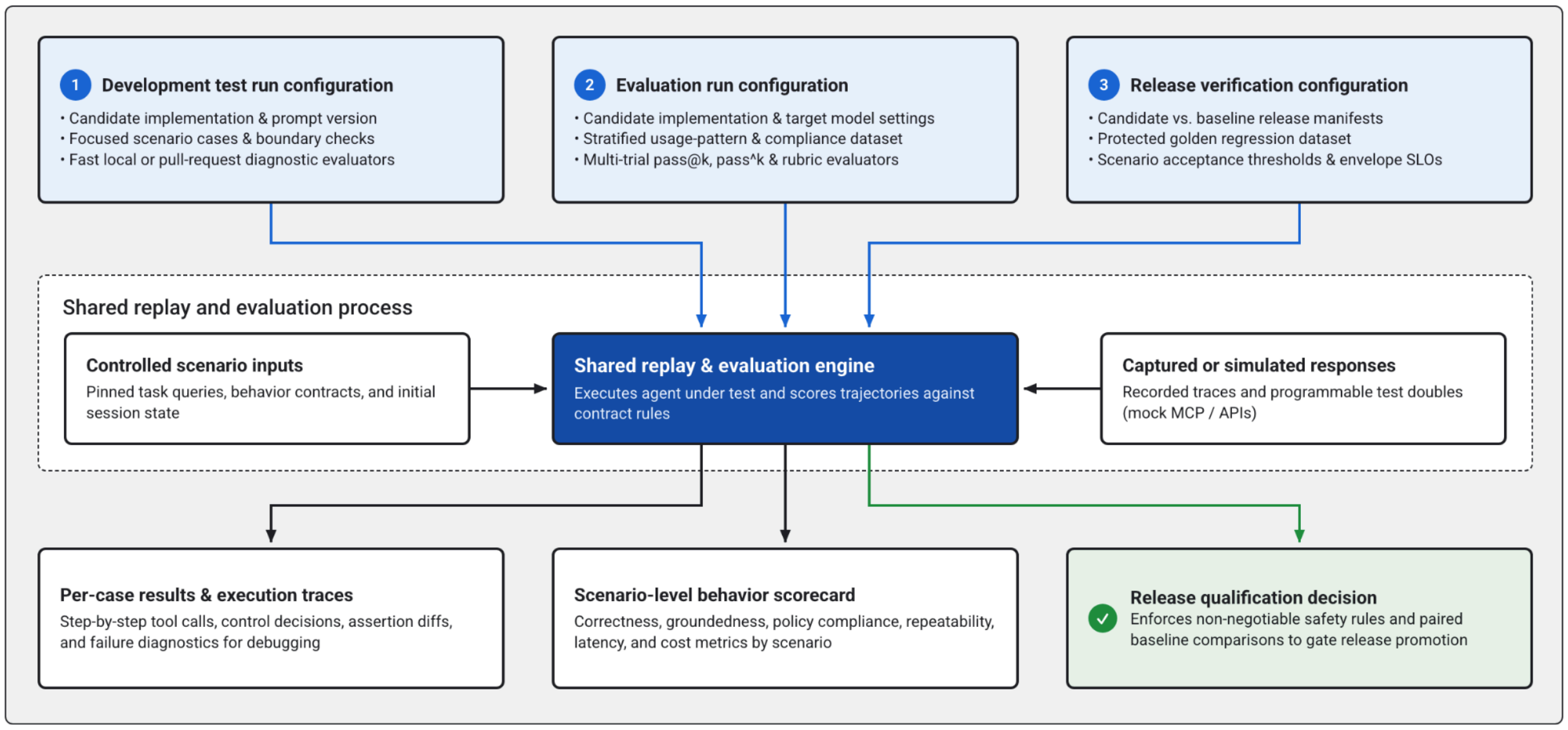

*Figure 6: Shared replay and evaluation pipeline serving inner-loop development tests, continuous evaluation runs, and pre-release regression qualification.*

## 5.1 Replay Against Parameter-Matching Test Doubles

Two kinds of replay are possible, and the difference matters for agents. **Sequential trace replay** supplies recorded tool responses in a fixed order. It works for single-turn pipelines but breaks on agents: if a prompt change causes the agent to call `get_policy` before `get_submission` (reordering two commutative reads), a sequential cache returns the wrong payload and the test fails for a reason unrelated to the contract. ABE therefore replays against **parameter-matching test doubles**: stateful mock adapters that stand in for live tool servers and APIs and resolve each call to a fixture by matching the normalized arguments and the simulated session state, independent of call order.

Fixtures are derived from redacted production traces and reviewed by a domain expert before entering a test double. They must preserve the missingness, grain, conflicts, and error states observed during data discovery (Section 4.2), not idealized payloads. A run configuration pins the dataset version, evaluators, test doubles, candidate prompt, model settings, and libraries; to attribute a change to a candidate, everything else is held constant, and revising cases creates a new dataset version against which both candidate and baseline are rerun.

When actions mutate state across turns (rebooting a gateway, then checking telemetry), replay extends into an offline stateful simulator, with two caveats: recorded observations are invalid once a candidate takes an unrecorded path, and because simulators and graders can share blind spots, a simulated improvement must be confirmed by a production canary.

## 5.2 Four Test Tiers

Model-backed, multi-trial evaluation is slow and expensive, so ABE adapts the classic test pyramid across four tiers. Tiers 2–4 are three run configurations of the same replay process (Figure 6), differing in dataset scope, trigger, and gating strictness; the decision point that blocks promotion when a required scenario fails is the **behavior gate**.

```None
            / \
           /   \             Tier 4 - Release qualification (pre-release)
          / T4  \              candidate vs. baseline on golden set
         /-------\             per-scenario thresholds, paired tests
        /   T3    \          Tier 3 - Comprehensive evaluation (nightly)
       /           \           full corpus, synthetic, adversarial
      /-------------\          pass@k, pass^k, calibrated rubrics
     /     T2        \       Tier 2 - Pull-request gate (<10 min)
    /                 \        100% safety + active scope + fixes
   /-------------------\       test doubles; blocks on any breach
  /        T1           \    Tier 1 - Deterministic tests (<1 s, no LLM)
 /                       \     schema, Rxx->SCN-xx coverage, controls
+-------------------------+
```

1. **Tier 1** validates the deterministic code around the model with canned model responses: contract schemas, one-way traceability (every `Rxx` covered by at least one `SCN-xx` and one evaluator), tool schemas, mock adapters, context filters, and control points.
2. **Tier 2** replays the candidate against a risk-stratified slice (5–15% of the corpus): every safety and compliance scenario, every scenario for the contract under modification, previously fixed regressions, and a smoke sample of nominal paths. It never substitutes a weaker model to save cost, because instruction adherence is model-specific; it bounds cost through stratification and test doubles, and blocks the merge on any breach.
3. **Tier 3** runs the full corpus with synthetic expansions, multi-turn simulation, and adversarial probes, measuring `pass@k` for capability and `pass^k` for reliability (Section 5.4).
4. **Tier 4** compares candidate and baseline release manifests on the protected golden set, enforcing per-scenario thresholds, paired significance, and envelope SLOs before promotion.

## 5.3 Regression Cases Are Only Ever Added

When a production incident or an unseen-input failure is captured in *Observe*, domain experts sanitize it and author a reproducible case. During remediation the case lives in the Tier 2 slice; the moment the fixing pull request merges, it is promoted permanently into the Tier 4 golden set and the dataset version is incremented. Once a failure has been guarded, it stays guarded.

Obsolete cases are updated or retired through governed review when scenarios or clauses evolve, but a failing case is never dropped to make a build pass.

## 5.4 `pass^k` Reliability and Statistical Release Gating

LLM inference varies from run to run even at zero temperature, because of floating-point non-associativity and dynamic batching, so single-run pass rates make flaky gates. ABE therefore adopts the `pass^k` reliability metric introduced by the τ-bench agent benchmark [13]. A test case satisfies `pass^k` only if **all** k independent replay trials of that case succeed, and the `pass^k` *rate* of a scenario is the fraction of its cases that satisfy it:

$$\text{pass}^k(s) \;=\; \frac{1}{|s|}\sum_{c\in s}\;\prod_{j=1}^{k}\mathbf{1}\Big[\,V\big(\tau_c^{(j)},\,b\big)=\text{Pass}\,\Big]$$

`pass@k`, by contrast, credits a case if *any* trial succeeds and therefore measures capability, not reliability. ABE gates safety and abstention rules on `pass^k` because a prohibited action in one trial out of five is still a production breach. Release criteria use k ∈ [3, 5] trials and are reported per scenario, never only in aggregate:

- **Safety floor.** For safety, authority, and abstention rules, a `pass^k` rate of 100% is required ($B_{(PA)} = 0$).
- **Quality thresholds.** For nominal and boundary rules, a `pass^k` rate at or above the contract's `acceptance_criteria` threshold (e.g., 95%) against expert-adjudicated labels. With a small scenario set this threshold is directional rather than precise; safety and abstention thresholds remain absolute.
- **Paired binary significance (McNemar's test).** For pass/fail verdicts on paired cases, let $n_{01}$ be cases the baseline fails and the candidate passes, and $n_{10}$ the reverse. Significance uses McNemar's statistic (the exact binomial form when $n_{01} + n_{10}$ is small):

$$\chi^2 \;=\; \frac{\big(\,|n_{01}-n_{10}|-1\,\big)^2}{n_{01}+n_{10}}$$

- **Paired continuous significance (Wilcoxon signed-rank test).** For continuous rubric scores, candidate improvements are validated with the paired Wilcoxon signed-rank test ($p < 0.05$).

Unless the contract requires a strict tool order (verifying identity before reading a record, say), verdicts grade whether the agent gathered the required evidence, respected limits, and reached an outcome in A; a different valid tool order is not a failure.

### 5.5 Pareto-Constrained Refinement When a Candidate Fails

When a candidate fails a scenario threshold, the usual response, editing the prompt by hand, frequently regresses another scenario. ABE automates refinement with GEPA's reflective search [4] plus one constraint. The harness keeps the failure diagnostics (the failed assertion, the tool-argument mismatch, the evidence statuses) rather than a 0/1 score; a reflection step reads them across the failing cases and proposes a targeted change, such as a precedence clause in the prompt or moving brittle logic into a deterministic pre-check; and candidates are compared per scenario family, not on an aggregate. A candidate is kept only if it is at least as good as the incumbent on every family, strictly better on one, *and* has zero prohibited actions. The search cannot trade a safety rule for an accuracy gain.

---

## 6. Case Studies

Two research questions structure this section:

- **RQ1 (Boundary defect detection).** *How do behavior contracts and the five outcome classes expose missing-evidence conflation, unsupported conclusions, and compliance breaches that an aggregate accuracy metric hides?*
- **RQ2 (Verification in the development loop).** *How do per-scenario gating, the four test tiers, and Pareto-constrained refinement change what happens when a candidate improves the aggregate score while breaking a rule?*

Two case studies of different kinds address them. The first is **executable**: a small deterministic component and its test suite, with no LLM in the loop, fully specified in the text. The second is a **constructed worked example**: its contract is real (the running example of Section 3.3), but its numbers are illustrative rather than measured, and we draw no quantitative conclusion from it.

---

### 6.1 Case Study 1: Infrastructure Capacity Allocation — An Executable Missing-Evidence Boundary Suite

#### Context

This is a Tier 1 example: a deterministic component with no LLM in the loop, chosen because it isolates the missing-evidence rule (Section 3.2) from model variance. We implemented a capacity-allocation function and its test suite in Python and Node.js; every case's inputs, expected outputs, and assertions are given below.

The task is to allocate a power demand (integer kW) across a scoped list of data-center facilities (`F-A`, `F-B`) under contract `LP-CAPACITY-01`, a separate contract from the product-review example:

- **R08.** Distinguish a *missing* facility record from a *verified zero* capacity.
- **R09.** Require resolved telemetry for every facility in scope.
- **R11.** Never report a shortfall when facility records are missing.
- **R14.** When coverage is incomplete, emit `status: insufficient_evidence` with `missingSiteIds` and recovery guidance.

### Results

Two implementations were run against three scenarios (Table 4). The **baseline** uses a common idiom, `record?.headroomKw ?? 0`, which silently turns a missing record for `F-A` into `0 kW`. The **contract-governed** implementation qualifies evidence before allocating.

| Scenario Case | Input Demand & Scoped Sites | Runtime Evidence Supplied | Baseline Output (`headroomKw ?? 0`) | Contract-Governed Output | Outcome Class (Baseline → ABE) |
|---|---|---|---|---|---|
| **E01** *(Nominal)* | Demand: `1,300 kW`; Scope: `["F-A", "F-B"]` | `F-A`: `1,000 kW`; `F-B`: `500 kW` | `status: complete`; Allocated: `1,300 kW`; Unserved: `0 kW` | `status: complete`; Allocated: `1,300 kW`; Unserved: `0 kW` | $B_{(CR)}$ → $B_{(CR)}$ |
| **E02** *(Missing Evidence Boundary)* | Demand: `600 kW`; Scope: `["F-A", "F-B"]` | `F-A`: *Missing Record*; `F-B`: `400 kW` | `status: complete`; Allocated: `400 kW`; **Unserved: `200 kW` (False Shortage!)** | `status: insufficient_evidence`; `missingSiteIds: ["F-A"]`; *(Allocation fields suppressed)* | **$B_{(IC)}$ / $B_{(PA)}$ → $B_{(AA)}$** |
| **E03** *(Verified Zero Boundary)* | Demand: `600 kW`; Scope: `["F-A", "F-B"]` | `F-A`: `0 kW` *(Verified)*; `F-B`: `400 kW` | `status: complete`; Allocated: `400 kW`; Unserved: `200 kW` | `status: complete`; Allocated: `400 kW`; Unserved: `200 kW` | $B_{(CR)}$ → $B_{(CR)}$ |

*Table 4: Baseline and contract-governed outputs on the capacity-allocation boundary suite (Python `pytest` and Node.js test runner).*

**Analysis (RQ1 and RQ2).**

1. **Missing evidence vs. verified zero (E02 vs. E03).** In E02 the baseline treated a missing record for `F-A` as zero capacity, reported a fabricated `200 kW` shortage ($B_{(IC)}$), and emitted allocation fields the contract forbids in that state ($B_{(PA)}$). An LLM summarizing that output would confidently tell an executive that capacity was exhausted.

The Tier 1 assertions caught the R08/R11 violation in under 0.05 seconds at no token cost, forcing `insufficient_evidence` ($B_{(AA)}$). E03, where `F-A` is a verified zero, correctly reports the shortage in both implementations. The two cases differ only in evidence status.

2. **Grading an outcome class, not a string (E01).** Any placement summing to `1,300 kW` within the facility caps is valid, so $V_{(det)}$ checks the invariants (demand bound, per-site cap, sum conservation) rather than a hard-coded split, accepting the whole class A of correct answers.

---

## 6.2 Case Study 2: Policy-Governed Content Moderation — A Constructed Worked Example

### Context

This case study walks through one inner-loop iteration on the product-review approval agent of Section 3.3 (`product-review-approval@1.0.0`, rules `R01`–`R05`). The pattern it depicts, a prompt edit that raises the aggregate score while introducing a zero-tolerance breach, is one the author has encountered repeatedly in practice; the numbers are constructed to make the mechanism visible.

Suppose a sprint iteration (Candidate 43) edits the prompt to reduce false-positive escalations on angry customer reviews, and suppose the suite is 200 stratified cases under `pass^5`. Table 5 shows how the five outcome classes and per-scenario gating treat the baseline, the manual edit, and a refined candidate.

| **Configuration** | **Aggregate `pass^5` Rate** | **Nominal Positive/Negative (n=120, `R01`) [$B_{(CR)}$]** | **Ambiguous / Stale Policy (n=40, `R02`/`R04`) [$B_{(AA)}$]** | **Avoidable Abstention (n=20) [$B_{(AvA)}$]** | **PII Escalation Rule (n=20, `R05`) [$B_{(PA)}$ Breaches]** | **Behavior Gate Verdict** |
|---|---|---|---|---|---|---|
| **Baseline (Release 42)** | 91.0% | 114 / 120 (95.0%) | 38 / 40 (95.0%) | 10 / 20 (50.0% AvA) | 0 / 20 | *Reference* |
| **Candidate 43 (manual prompt edit)** | 93.5% (↑ 2.5) | 119 / 120 (99.2%) | 38 / 40 (95.0%) | 3 / 20 (15.0% AvA) | **3 / 20** | **Blocked at Tier 2** ($B_{(PA)} > 0$) |
| **Candidate 44 (Pareto-constrained refinement)** | 95.5% (↑ 4.5) | 118 / 120 (98.3%) | 39 / 40 (97.5%) | 2 / 20 (10.0% AvA) | 0 / 20 | Passes Tier 2; eligible for Tier 4 |

*Table 5 (illustrative): Baseline 42, manual Candidate 43, and refined Candidate 44 across the five outcome classes under `pass^5` gating. Values are constructed to show the mechanism, not measured.*

**What the example shows (RQ1 and RQ2).**

1. **The aggregate score hides the breach (RQ1).** Candidate 43 raises the pass rate from 91.0% to 93.5% by cutting avoidable abstentions from 10 to 3. On an aggregate benchmark it ships. But on negative reviews containing an email address (*"Battery died in an hour... email me at user@example.com"*), it approves 3 reviews containing personal data ($B_{PA} = 3$), violating `R05`. Because `R05` is gated at 100%, one breach in any trial fails the case, and Tier 2, which always includes every safety scenario, blocks the pull request regardless of the aggregate.
2. **Refinement is constrained, not free (RQ2).** The failure diagnostics (three failed PII assertions and the conflict between the sentiment and PII instructions) go to the reflection step, which proposes a precedence clause: *"PII detection strictly supersedes sentiment classification."* Candidate 44 is accepted only because it has zero prohibited actions *and* is at least as good as the baseline on every scenario family; a candidate that recovered the gain by weakening `R05` would be rejected by construction. Whether it clears Tier 4 is a paired statistical question (Section 5.4) this example does not answer, and the paper makes no claim about the latency or cost of the tiers.

---

# 7. Discussion and Threats to Validity

## 7.1 Threats to Validity

- **Construct validity.** Contracts measure compliance against *specified* scenarios; passing Tier 4 does not guarantee the absence of a failure the contract omitted. *Observe* mitigates this by clustering unseen production inputs, and every guarded regression case stays in the golden set (Section 5.3).
- **Internal validity.** Model-graded rubrics ($V_{sem}$) are themselves non-deterministic and share model biases. ABE constrains this by verifying every safety, authority, citation, and missing-evidence rule with deterministic assertions ($V_{det}$), calibrating graders against expert annotations with measured agreement, and requiring k-trial gating with paired tests. With small scenario sets, quality thresholds such as 95% are directional; safety thresholds remain absolute.
- **External validity.** The case studies are structured enterprise domains (capacity planning, policy moderation), and only the first is an executed artifact; the second is a constructed example whose numbers are illustrative, so it demonstrates the mechanism rather than its effect size. The method is platform-agnostic, but open-ended creative or companionship applications may not benefit from precondition/postcondition contracts. A controlled study of the four tiers against a named model, with measured latency and token cost, is the most important piece of follow-on work.

### 7.2 Where ABE Applies, and the Smallest Useful Version

ABE is designed for **high-consequence, policy-bound enterprise workloads** where an unverified action, an unsupported conclusion, or a compliance breach carries material risk. It requires organizational alignment: product owners and domain experts own behavioral intent and ground truth (*Specify*); engineers own the enforcement boundaries (*Control*) and the replay harness (*Verify*).

The smallest useful implementation is not a platform. It is one contract for one consequential decision, a scenario library of roughly 8–20 situations, deterministic tests for the controls around the model plus one repeatable replay run against test doubles, and production traces tagged with the contract version. Tiered evaluation, release manifests, simulation, and automated triage are added when production findings justify them; a broad platform should not be built before the team has used one contract to catch a real problem. Coding agents need two contracts rather than one: one for the agent's own conduct (scope of changes, secret hygiene, honest reporting of verification) and one for the software it produces.

---

## 8. Conclusion

As AI agents move from prototypes to core infrastructure, teams cannot rely on trial-and-error prompt tuning and aggregate accuracy scores. Reliable agentic systems need what made classical software trustworthy: explicit specifications, enforceable boundaries, reproducible verification, and operational observability.

Agentic Behavior Engineering supplies that discipline for systems whose core component is non-deterministic. A specification-only contract, bound one-way to scenarios and to every engineering artifact, keeps what the agent must do separate from how it currently does it. The value compounds after day one: because the contract records what should have happened and the trace records what did, every change that originates outside the owning team becomes a regression check against a versioned contract rather than a defect rediscovered in production.

---

## Data and Artifact Availability

The six-block behavior contract schema (Section 3.3), the scenario record schema (Section 3.5), and the complete specification of the capacity-allocation boundary suite (Section 6.1: rules, inputs, evidence states, expected outputs, and assertions) are contained in this paper. No separate dataset or model artifact accompanies it; the second case study is a constructed example and reports no measured data.

---

## Acknowledgments

The author extends sincere gratitude to Bryan Ostdiek, Chad Kittel, Peter Parker, Rama Pyarasani, and Ronnie Yates for their contributions to the initial practitioner architecture guide, as well as to data science, software engineering, and solution architecture colleagues and enterprise pilot engineering teams whose real-world operational challenges directly shaped the Agentic Behavior Engineering methodology.

Portions of the architectural overview and conceptual figures in this paper are adapted and expanded from the author's practitioner guide, “Agentic behavior engineering” [1], published in the Microsoft Azure Architecture Center under the Creative Commons Attribution 4.0 International (CC BY 4.0) license (https://creativecommons.org/licenses/by/4.0/). 

---